\documentclass[showpacs,aps,floatfix,showkeys]{revtex4}
\usepackage{graphics}
\usepackage{bm}

\begin{document}

\title{Weighted and unweighted network of amino acids within protein}
\author{Md. Aftabuddin} 
\author{Sudip Kundu}
\email{skbmbg@caluniv.ac.in}
\affiliation {Department of Biophysics, Molecular Biology \& Genetics\\
University of Calcutta, 92, A.P.C. Road, Kolkata- 700009, India}

\date{}

\begin{abstract}
{The information regarding the structure of a single protein is encoded
in the network of interacting amino acids. Considering each protein as a 
weighted and unweighted network of amino acids we have analyzed a total of
forty nine protein structures that covers the three branches of life on earth.
 Our results show that the probability degree distribution of network 
connectivity follows Poisson's distribution; whereas the probability 
strength distribution does not follow any known distribution. 
However, the average strength of amino acid node depends on its degree (k).
For some of the proteins, the strength of a node increases linearly with 
k. On the other hand, for a set of other proteins, although the
strength increases linaerly with k for smaller values of k, 
we have not obtained any clear functional relationship of 
strength with degree at higher values of k. The results also show that 
the weight of the 
amino acid nodes belonging to the highly connected nodes tend to have
a higher value. The result that the average clustering coefficient of
weighted network is less than that of unweighted network implies that
the topological clustering is generated by edges with low weights. 
The ratio of average clustering coefficients of protein network to
that of the corresponding classical random network varies linearly
with the number (N) of amino acids of a protein; whereas the ratio 
of characteristic path lengths varies logarithmically with N. The
power law behaviour of clustering coefficients of weighted and 
unweighted network as a function of degree k indicates that the network
has a signature of hierarchical network. It has also been observed that
the network is of assortative type. }
\end{abstract}
\maketitle

\section{Introduction }

A large number of researchers has been attracted to shed light on the topology, 
growth and dynamics of different kind of networks since 1990s. 
Studies on a large number of networks such as food webs, 
electrical power grids, the world wide web (WWW), coauthorship and citation 
network of scientists have been accelerated after the major breakthrough 
of Barabasi-Albert's work on the mapping
of WWW and the physical internet in the late 1990s [1-6]. 
Like all branches of science, 
network analysis is increasingly recognised as a powerful approach to understanding 
of even the biological organisation and the 
function of cellular components and may also help us to understand the principles 
driving the evolution of
 living organism [7]. One could think the civilisation as a network of human and
 ecosystems. Whereas
ecosystems are interaction of different organisms. Multicellular organism
 is a network of unicellular organisms. Even the cell is nowadays considered as 
a outcome of interactions of complex networks of genome, transcriptome, proteome, 
metabolome. Efforts have also been made to study the protein-protein 
interaction network [8-12] and also the amino acid network within protein [13-17]. 
Network of amino acids
 within a protein shows the `small world' property, i.e, 
it is possible to connect 
any two amino acids of a protein through just a few links. It has also been
observed that long range interactions of amino acids have scale free
properties, i.e, they have a distribution of connectivities that decays 
with a power law tail. The scale free networks emerge in the context of 
growing network in which new vertices connect preferentially to the more
highly connected vertices in the network. On the other hand, the short
range interactions follow  a Poisson like distribution, a characteristic
of so called `random network' which emerges as the result of random link
within vertices. Recently, we have studied the networks of hydrophobic 
and hydrophilic amino acids within a protein separately [17] and have observed the 
small world properties for both the cases. Our studies also show that 
the average degree of the hydrophobic node is larger than the average degree of 
hydrophilic node. In all these studies protein has been considered as 
an unweighted network. Very recently, Barat {\it et al} [18] have proposed a 
framework of understanding the 
weighted network and have analysed the traffic and coauthorship networks.
 Amino acid network within a protein can be one of the best examples of the
weighted network. In this paper, 
we have applied the concepts of weighted network to analyze and understand 
 the native protein structures and their different network properties.

\section{Methods } 

\subsection{Protein as a weighted network}

Protein molecule is a polymer of twenty different amino acids 
joined by peptide bonds. The selection of specific amino acids 
is crucial for the molecule's native structure. Amino acids in 
different regions form 
local regular secondary structure ($\alpha$ - helix, $\beta$-
sheet, etc.), which pack to form tertiary structure (active conformation) 
that in turn stabilizes three dimensional conformations. 
Several polypeptide chains forming the tertiary structure 
arrange to form quarternary structure (homo/hetero dimer complexes) [19]. 
In active 3-D conformational space of a protein, 
the amino acids may have different interactions with 
each other. In fact, the i-th amino acid of a protein may come in 
close contact with (i+j)-th amino acid, where j may take a large value.
Further, between any two amino acids  
the atoms of the side chains of two different amino acids may have 
the possibilities of more than one interaction. 
In the amino acid network within a protein,
the amino acids are considered as  nodes. A link between any 
two amino acids is considered to be present if any two atoms from two
different amino acids are within 5 \AA~ distance 
(which is the higher cut-off of 
London-van der Waals forces [20]). Since several atoms of any amino acid `i' may be 
within the prescribed distance of several atoms of another amino acid `j'; 
there is a possibility of 
multiple links between different amino acids. 
These multiple links between any two different amino acids are
  the basis of the weight of the connectivity. 
The intensity $w_{ij}$ of the interaction 
between two amino acids `i' and `j' is defined as the number 
of possible links between i and j-th amino acids.
To understand the nature of link we have analysed 
a total of forty-nine protein structures 
[Table I]  of nine different fold types 
that almost covers all the major four protein 
classes (all -$\alpha$,
 all -$\beta$, mixed $\alpha / \beta$, mixed $\alpha + \beta $) 
corresponding to the three 
branches of life on earth (archaea, bacteria, eukariya) [21].

\subsection{Network Parameters}

The most elementary property of the connectivity of a network is 
the degere of its node. The degree of any node  
`i' is represented by $k_i = \sum_{j} a_{ij}$. Here $a_{ij}$ is 
the element of the adjacency 
matrix of the graph, whose value is 1 if an edge 
connects a node `i' to the another 
node `j' and 0 otherwise. For a weighted network, one wants to 
calculate the strength of a node represented by 
$ s_i = \sum_{j} a_{ij} w_{ij} $. Here 
$w_{ij}$ is the number of possible interactions between any 
two (`i' and `j' ) amino acids.
This parameter actually is a better representation of a protein 
network since it actually represents the number of connectivity 
of an amino acid with all other amino acids. It should be mentioned 
that we have not considered the  energy of 
any interaction. The spread in the strength of a node has been
 characterised by a distribution function $P(s)$;
 where $P(s) = N(s)/ \sum N(s)$, $N(s)$
 being the number of nodes with strength `s'. On the other hand 
the probability of degree distribution is represented 
by $P(k) = N(k)/ \sum N(k)$. 

	 To examine if there is any 'small world' property in the 
network, one conventionally have to determine the two 
parameters- i) the characteristic path length ($L$) and 
ii) the clustering coefficient ($C$).  The characteristic path 
length $ L$ of a graph is the path length between 
two nodes averaged over all pairs of nodes.  Traditionally  
the clustering coefficient $ C_i$ of a node 'i' is 
the ratio between the total number $ (e_i) $ of 
the edges actually connecting its nearest neighbour 
and the total number of all possible 
edges between all these nearest neighbour 
[$k_i ( k_i - 1) /2$ ; if  'i' vertex has $k_i$ 
neighbours] and is given by $C_i = 2 e_i / k_i ( k_i - 1)$. 
The clustering coefficient of the whole network is the average 
of all individual $C_i$'s. For a random 
network having N number of nodes with average degree $<k>$, 
the characteristic pathlength ( $L_{r} $ ) and the clustering 
coefficient ( $C_{r}$ ) have been calculated using the expression  
$L_{r} \approx $ lnN/ln$<k>$  and $C_{r} \approx$ $<k>$ / N  
given in [2]. According to Watts \& Strogatz [2], 
if $ L$ and $C$ values of a network are such that $ C >> C_{r} $ and
$ L \ge L_{r} $ ; then that network is said to have the 'small world' property. 

Combining the topological information with the weight distribution 
of the network, Barat et al have introduced an analogous parameter
of C and that is known as weighted clustering coefficient, $C_{i}^{w}$.
The weighted clustering coefficient is given by
\begin{displaymath}
C_{i}^w = \frac{1}{s_i(k_i -1)} \sum_{j,h} \frac{w_{ij} +w_{ih}}{2} a_{ij} a_{ih} a_{jh}
\end{displaymath}
This weighted clustering coefficient, $C_{i}^w$
is a measure of the local cohessiveness that takes into account 
the importance of the clustered structure on the basis of amount of 
interaction intensity (number of possible interactions between amino acids)
actually found on the local triplets. 

	To study the tendency for nodes in networks to be connected 
to other nodes that are like (or unlike) them,  we have 
first calculated  the Pearson correlation coefficient (r) of
 the degrees at either ends of an edge. For our undirected unweighted 
protein network this r value has been calculated using the expression 
suggested by Newman [21] and is given as
\begin{displaymath}
r = \frac{N^{-1} \sum_i j_i k_i - [N^{-1} \sum_i 0.5 (j_i + k_i)]^2}{N^{-1} \sum_i 0.5 (j_i^2 + k_i^2) - [N^{-1} \sum_i 0.5(j_i + k_i)]^2}
\end{displaymath}

Here $j_i$ and $k_i$ are the degrees of the vertices at the
ends of the i-th degree, with i=1,..N. 
This parameter would help us to understand whether the network is 
assortative or disassortative type. For an unweighted  network 
this could also be understood calculating the average nearest neighbor 
degree for different $k$ and comparing this $k_{nn}$ with the $k$ values.
If $k_{nn}(k)$ is an increasing function of $k$ then the network is 
of assortative type. On the other hand, for a weigheted network
Barat et al have modified the expression for $k_{nn,i}$ and 
suggested the equivalent weighted average nearest-neighbors 
degree $k_{nn,i}^w$ to be defined as 
\begin{displaymath}
k_{nn,i}^w = \frac {1}{s_i} \sum_{j=1}^N a_{ij}w_{ij}k_{j}
\end{displaymath}
 
\section{Results \& Discussions}
   We have analyzed the network of forty-nine proteins listed in Table I. 
For each of all the proteins, all the amino acids are part of a single network, i.e., there is no isolated node present in any of the networks.    
For each of the 49 proteins we have calculated the average degree $ <k>$.
The values vary from 8.23 to 11.01 . The average of the $<k>$ values 
for all 49 proteins was found to be 9.40 with  standard deviation 0.55. 
We have also observed that the value of average degree does not depend on
the network size (i.e., on the number of amino acids of the protein). 
Next we have studied the degree and strength distributions of the nodes within
the network. For all the proteins we have studied,
the probability distributions $P(k)$  are of Poisson's distribution 
type as is evident from Fig 1. However, for each of  the proteins the probability 
strength distribution $P(s)$ of the nodes exhibit no clear 
pattern as is shown in Fig 2.
 This dissimilarity 
in the functional behaviour of $P(k)$ and $P(s)$ demands further
investigations on the relationship between strength($s$) 
and degree($k$) of the nodes. 

To understand the relationship between the strength of a node 
with its degree we have 
further studied the average strength $s(k)$ as a function of $k$. 
We have observed that the strength of a vertex changes with its 
degree $k$ [Fig 3]. For some of the proteins, there is a linear relationship 
of $s(k)$ with $k$ ,i.e., 
the strength of a vertex is simply proportional to its degree.
On the other hand, for a set of other proteins, although the
strength increases linearly with k for smaller values of k, we have not
obtained any clear functional relationship of strength with degree 
at higher values of k. 
One of the possible explanations is as follows.
Consider an amino acid `i' having a small number of  
amino acids interacting with it [Fig 4]. 
In this case, it is possible to accomodate most of the atoms of side
chains of all interacting amino acids within  5 \AA~ distance of
the i-th amino acid. Hence, for smaller k, the strength of a vertex
increases almost linearly with its degree. 
On the other hand, for larger k, there are more 
number of interacting amino acids than in the previous case. In such case, 
most of the atoms of a section of interacting amino acids may not fall
within the prescribed distance due to spatial constraints
imposed by the surrounding amino acids. Next, we have plotted the 
average weight ($<w_{ij}>$) as a function of the end-point degree 
($k_{i}k_{j}$) in Fig 5. The average weight fluctuates  
for the whole range of $k_{i}k_{j}$ values. 
 However, from Fig 5, it is clear that
although there are local fluctuations, there is a tendency
of the average weights to increase with their $k_{i}k_{j}$ values 
(as depicted from the inset figure of Fig 5). The result implies that
the weight of amino acid nodes belonging to the highly connected nodes
tend to have a higher value.

   We next describe how the clustering 
coefficients for both weighted and unweighted networks vary 
with their degree k. At the 
same time we examine whether the networks have the 
`small world' property or not.
We also test whether there is any dependency of average clustering 
coefficient and average path length on the network sizes. 

As mentioned earlier, we have followed Watts and Strogatz prescription 
to verify whether the protein network exhibit 'small world'
 property or not. 
We have calculated the average 
clustering coefficient $<C>$ and the characteristic path length $<L>$ for
each of the forty-nine proteins 
and their respective values ($<C_{r} >$ \& $<L_r>$) for the random
network having the same N and $<k>$.
 In the present study, the $<C>$ values vary from 0.464 to 0.586; whereas 
the ratios ($p = <C> / <C_r>$) of average clustering coefficient to 
that of the corresponding classical random graph vary from 4.61 to 25.20. 
On the other hand, the characteristic path 
length  is of the same order as that of the corresponding 
random network. Although the ratios (p) for protein networks are
 not of the order of $10^2 - 10^4$ as observed in the case of 
scientific collboration networks and network of film 
actors, there are several other networks where `p' may have  
smaller values [1, 4-6, 17]. For example, the ratio for metabolic
network, protein-protein interaction network, food webs and network 
of {\it C.Elegans} has values 5.0, 4.4, 12.0 and 5.6 respectively.
The above results indicate the small world property of the 
protein network.

From the above examples, it is clear that the ratio 
(p) varies from a smaller value to a very large value.  In our 
protein networks, the ratio (p) also varies in a wide range. 
Moreover it is also observed that the ratios (q) of $<L>$ and $<L_r>$
are not constant. We have further studied 
the dependencies of p and q on N [Fig 6].
It has been observed that both the ratios p and q  
vary with the number of nodes (i.e., the number of amino acids);
 but with a different relationships. The ratio (p) of clustering 
coefficients varies linearly with the number (N) of amino acids of 
the protein; whereas the ratio (q) of characteristic path lengths
varies logarithmically with N.

	We have already mentioned that the average clustering coefficient 
of the protein networks varies from 0.464 to 0.586. However, the average 
weighted clustering coefficient values for the same set of protein 
networks vary from 0.249 to 0.325.
The values of $<C>$ and $<C_w>$, when averaged over all of the
forty-nine proteins we have studied, are 
0.511 and 0.269 respectively [Table II]. 
And for each of the proteins, 
the $<C_{w}>$ value is always less than that of $<C>$. 
It implies that the topological clustering is generated by edges 
with low weight. The results also indicate that the largest part 
of interactions (i.e., interactions between two amino acids ) is occuring 
on edges (amino acids) not belonging to interconnected triplets. 
Therefore the clustering has a  minor effect in the organization of 
amino acid network within protein.
 
  Next we are interested to study if there is any hierarchy in the 
amino acid network. In a hierarchical network, the low degree nodes 
belong generally to well interconnected communities (high clustering 
coefficients) with hubs connect many nodes that are not directly 
connected (small clustering coefficient).
It has been stated that the signature of the hierarchical modularity 
lies in the scaling coefficient of $C(k) \sim k^{- \beta}$, when $\beta $ 
has a value of 1; whereas, for a non-hierarchical network the value of 
$\beta$ is 0 [6-7]. For the amino acid network within proteins,
both the $C(k)$ and $C^w(k)$ exhibit a power-law decay as a 
function of $k$ as is evident from Fig 7. 
The scaling coefficient ($\beta$) for the $C(k)$ varies from 0.247 to 0.367 
with an average of 0.322; whereas the corresponding coefficient ($\beta_1$) 
for $C^w(k)$  varies from 
0.525 to 0.722 with an average of 0.637  [Table II]. First of all, 
we get a power law decay for both C(k) and 
$C^w(k)$, but the values of the scaling coefficients lie within 
the range of 0 and 1. The values of the scaling coefficients imply that 
the networks have a tendency 
of hierarchical nature. Surprisingly, the values of $<C>$ are nearly double 
that of $<C_w>$, but the average value of $\beta$ is nearly half 
that of $\beta_1$. The result implies
that topological clustering is generated by amino acid edges with
low weights. Moreover, for smaller $k$, the amino acids within a
cluster are linked through higher weights than those related with 
larger $k$ values. For larger $k$ values, due to spatial constrains imposed by 
neighboring amino acids within the cluster, the nodes (amino acids) have less 
number of interactions between them. As k increases, the number of 
possible interactions (weights)  among amino acids within the cluster
decays faster than their connectivities [Fig 4].   

We have also calculated the Pearson Correlation coefficients (r) for 
each of the protein network. The r values vary from 0.18 to 0.38 with
 an average of 0.26 and standard deviation 0.05 [Table II].
The positive r values suggest the assortative mixing behaviour of 
the nodes of the network [21]. In a protein, the amino acids with high degree 
have a tendency to be attached with the amino acids having high k values. 
We have also calculated the average degree of nearest neighbors for both 
unweighted and weighted networks. For each of the proteins,
both the weighted and unweighted nearest neighbor values increase intially 
with increment of k, but for a larger k there is 
a tendency of saturation [Fig 8], which might be explained as the steric
hindrance of connecting amino acids due to three dimensional structural
organisation of the protein. Because of such steric hindrance, 
the position of any amino acid in 3D conformational space is restricted 
resulting in the maximum values of degree  ($k_{max}$) and strength 
($s_{max}$) of a node. In fact, all these network properties are 
mainly governed by two major important factors - (i) the movement 
of any i-th amino acid is restricted by (i-1)-th and (i+1)-th amino 
acids through peptide bonds; and (ii) the attachment of an amino acid 
to any other amino acid does not only depend on their physico-chemical
properties, but is also restricted by the spatial constraints imposed 
by neighboring amino acids, i.e., the preferential attachment is restricted 
by a maximum cut-off value due to steric hindrance. It should be mentioned 
that Amaral {\it et al} have discussed about the constraints 
limiting the addition of new links and argued that the nature of
such constraints may be the controlling factor for the emergence of different
classes of networks [23]. Hence, there is a need
of new model to understand the phenomenology behind the network of a 
polymer in three dimensional space.

\section{Summary}   

In summary, the amino acid network within a protein has a small world property. The probability 
degree distribution of the amino acid network connectivity within 
protein is of Poisson's distribution type; whereas 
the probability strength distribution does not follow any particular 
pattern. We have further observed that the strength of a node changes
 with its degree k. The ratios of the clustering coefficients of the protein network and its corresponding random network 
vary linearly with the number(N) of amino acids of the protein; 
while the respective ratios for characteristic path lengths vary 
logarithmically with N. The results further indicate that the topological 
clustering  is genereated by amino acids (edges) with low weights. 
The amino acid networks within protein exhibit the signature of a
 hierarchical network. The protein network is an outcome of assortative mixing of the amino acid nodes.

\begin{acknowledgements}
The authors acknowledge the computational support provided by 
Distributed Information Center of Calcutta University. The authors 
also thankful to Prof. U. Chaudhuri  and Dr. G. Gangopadhyay for their 
constant inspirations. 
\end{acknowledgements}

\newpage

\begin{table*}
\caption{List of PDB Ids of the proteins used in our study.}
\begin{tabular}{ccccccccccc}
\hline
1AVA\_c & 1CE7\_b & 1EYL\_a & 1HWN\_b & 1JLY\_b & 2AAI\_b & 1WBA & 7TIM\_a &1AGD\_b & 1BIH\_a \\
1CD8 & 1EPF\_d & 1HYX\_l & 1TLK & 3KBP\_c & 1HNG\_a & 1B8A\_a & 1E86\_a & 1FAP\_b & 1EOV\_a \\
1GVP & 1MJC & 1H7I\_a & 1JR8\_a & 1PSD\_a & 1RIS & 1URN\_c & 2ACY & 1VLT\_a & 1AYC\_a \\
256B\_a & 1CJ1\_a & 1FBV\_a & 1SHA\_a & 2HMZ\_a & 1BEB\_a & 1BJ7 & 1GM6\_a & 1OBP\_a & 1RPX\_a\\
1BMT\_a & 1DIO\_a & 1PDO & 3ECA\_a & 3RAB\_a & 1EUN\_a & 1G4T\_a & 1HO4\_a & 1NSJ & \\
\hline
\end{tabular}
\end{table*}

\begin{table}
\caption{The average clustering coefficient of weighted ($<C_{w}>$) and 
unweighted ($<C>$) network; the scaling coeffeicients of power law decay of 
clustering coeffients as a function of degree k and Pearson correlation coefficients of the protein networks.} 
\begin{center}
\begin{tabular}{ccccc}\hline
$ <C>$ & $ <C_{w}> $ & $C(k)$ $\sim$$ k^{-\beta}$&$ C_{w}(k)\sim k^{-\beta_{1}}$ & r\\
&&Value for $\beta$& Value for $\beta_{1}$&\\
\hline
0.511 $\pm$0.026 &  0.269$\pm$0.023 & 0.322$\pm$0.023&0.637$\pm$0.034&0.259$\pm$0.049 \\
\hline 
\end{tabular}
\end{center}
\end{table}

\newpage

\begin{figure}
\includegraphics{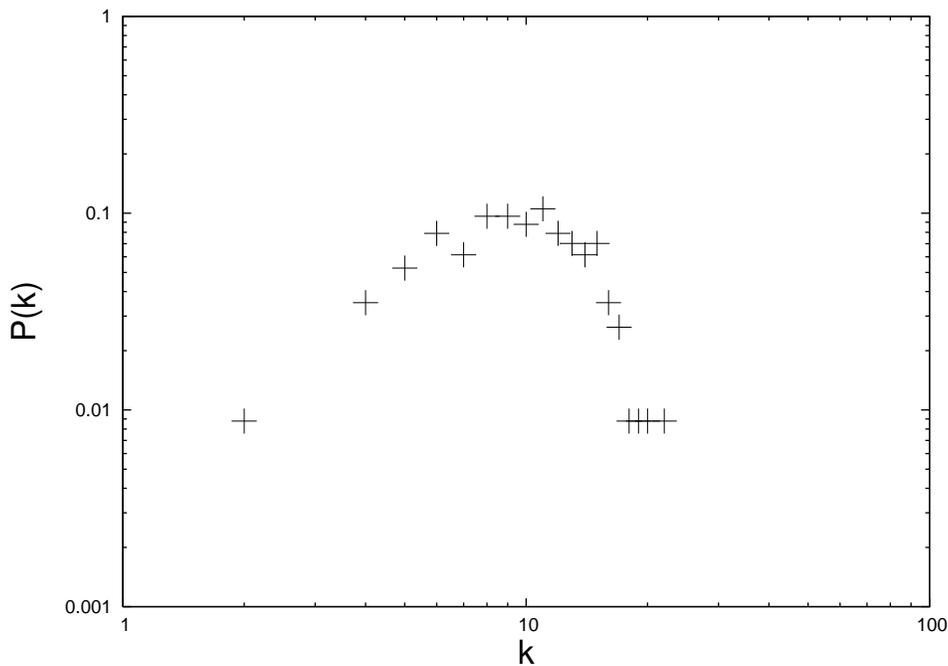}
\caption{A representative plot (PDB Id: 1CD8) of probability  degree 
distribution of amino acid network within protein. The degree k of any 
amino acid `i' is 
the number of amino acids interacting with i. The distribution follows 
Poisson's distribution.}
\end{figure}
\begin{figure}
\includegraphics{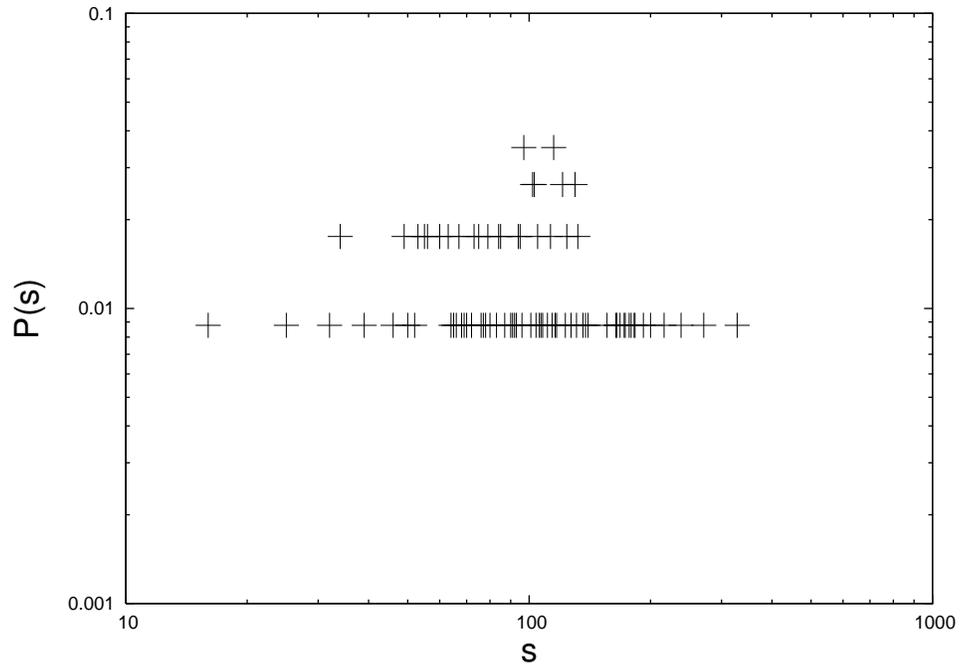}
\caption{A representative plot (PDB Id:1CD8) of probability strength 
distribution. The distribution has no definite functional form.}
\end{figure}
\begin{figure}
\includegraphics{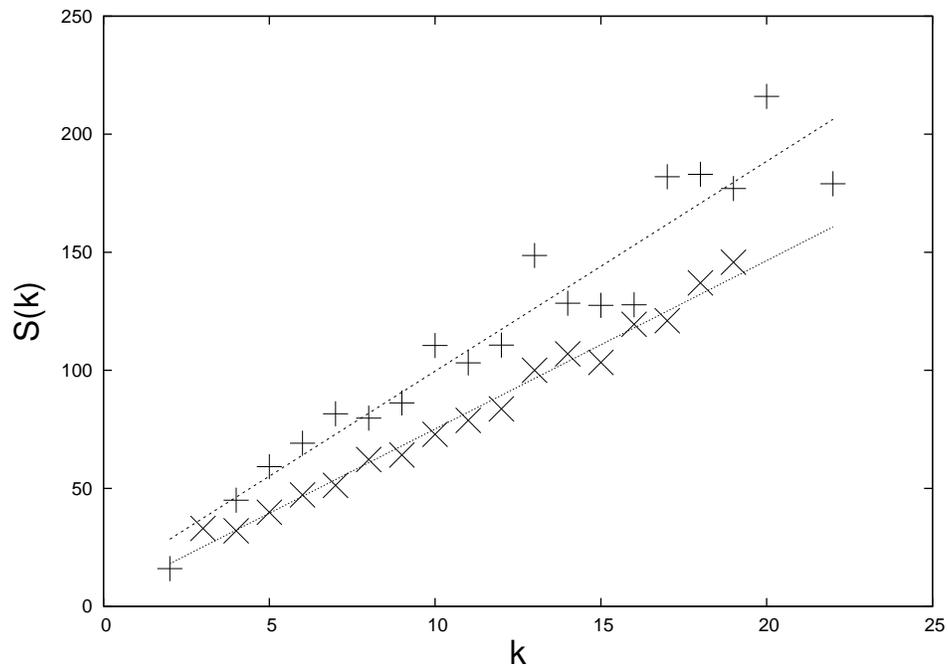}
\caption{A representative plot (PDB Id:1CD8, shown by plus sign 
and PDB Id:3KBP\_c, shown by cross sign) of average strength s(k) as function of
degree k of nodes. }
\end{figure}
\begin{figure}
\includegraphics{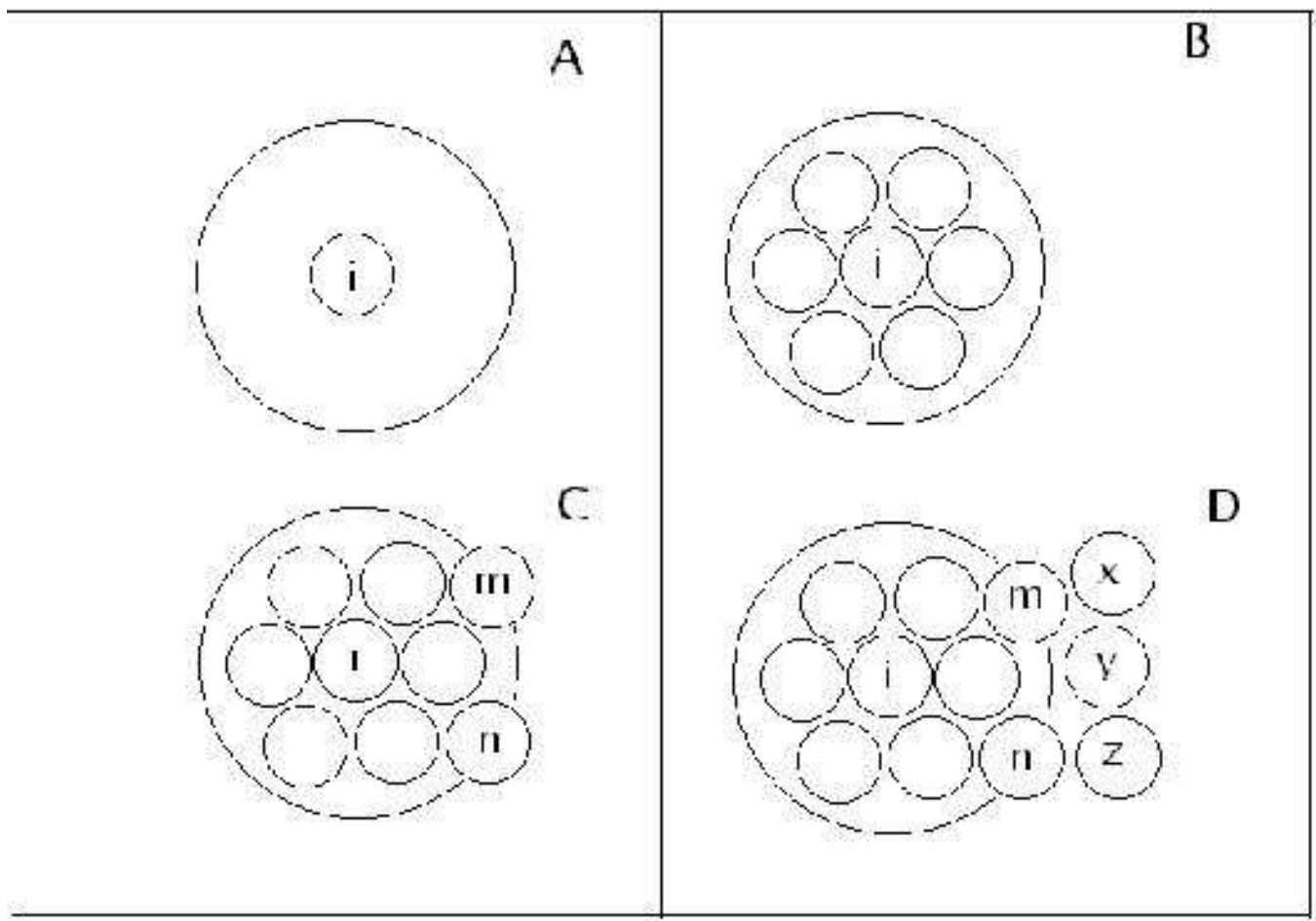}
\caption{Schematic diagram of (A) An amino acid `i' in 
conformational space. The large circle represents the 5 \AA~ distance from 
the line representing the i-th amino 
acid. Any amino acid within that large circle should have a possible interaction (link) with the i-th amino acid. (B) Atoms of all amino acids are within 
5 $\AA$ distance of i-th amino acid. (C) Some of the atoms of a section of
 amino acids (marked as `m' and `n') are not within that large circle. All 
the amino acids have a possible link with the i-th amino acid. All the atoms of
all amino acids, except two amino acids (marked as `m' and `n') have possible
interactions with i-th amino acid. However, a section of atoms (outside 
the large circle) of the two amino acids, `m' and `n', have no 
interaction with 
any of the atoms of i-th amino acid. (D) Three amino acids, `x',`y' and `z', 
are completely outside the large circle  and hence do not have any possible 
interaction with the i-th amino acid.}
\end{figure}
\begin{figure}
\includegraphics{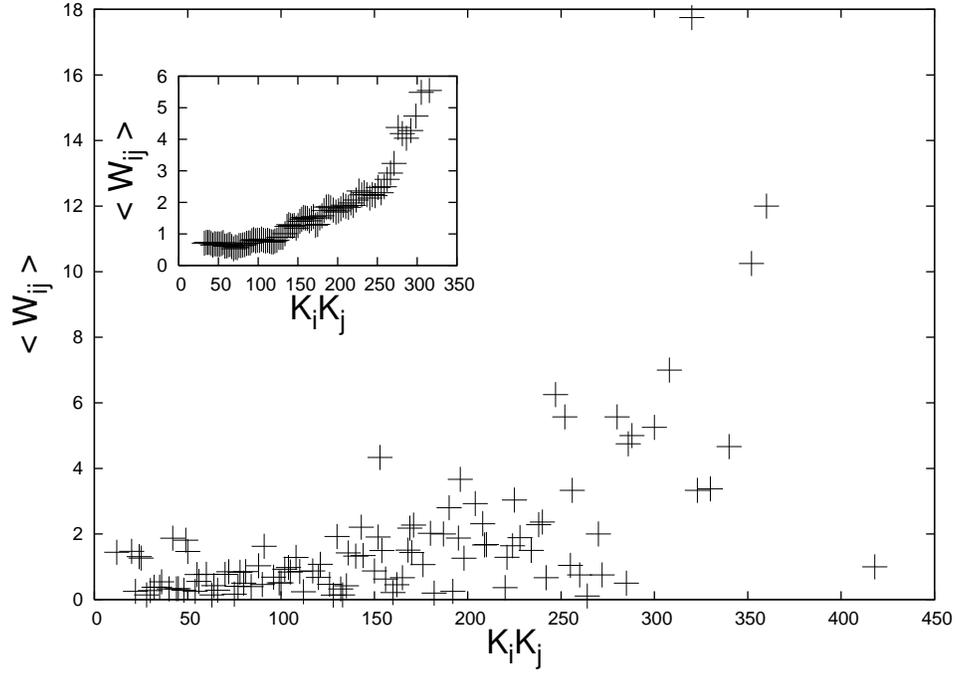}
\caption{A represntative figure (PDB Id:1CD8) of average weight as 
a function of the end-point degree. The running average of 
end-point degree with an window size of 15  is shown in inset.}
\end{figure}
\begin{figure}
\includegraphics{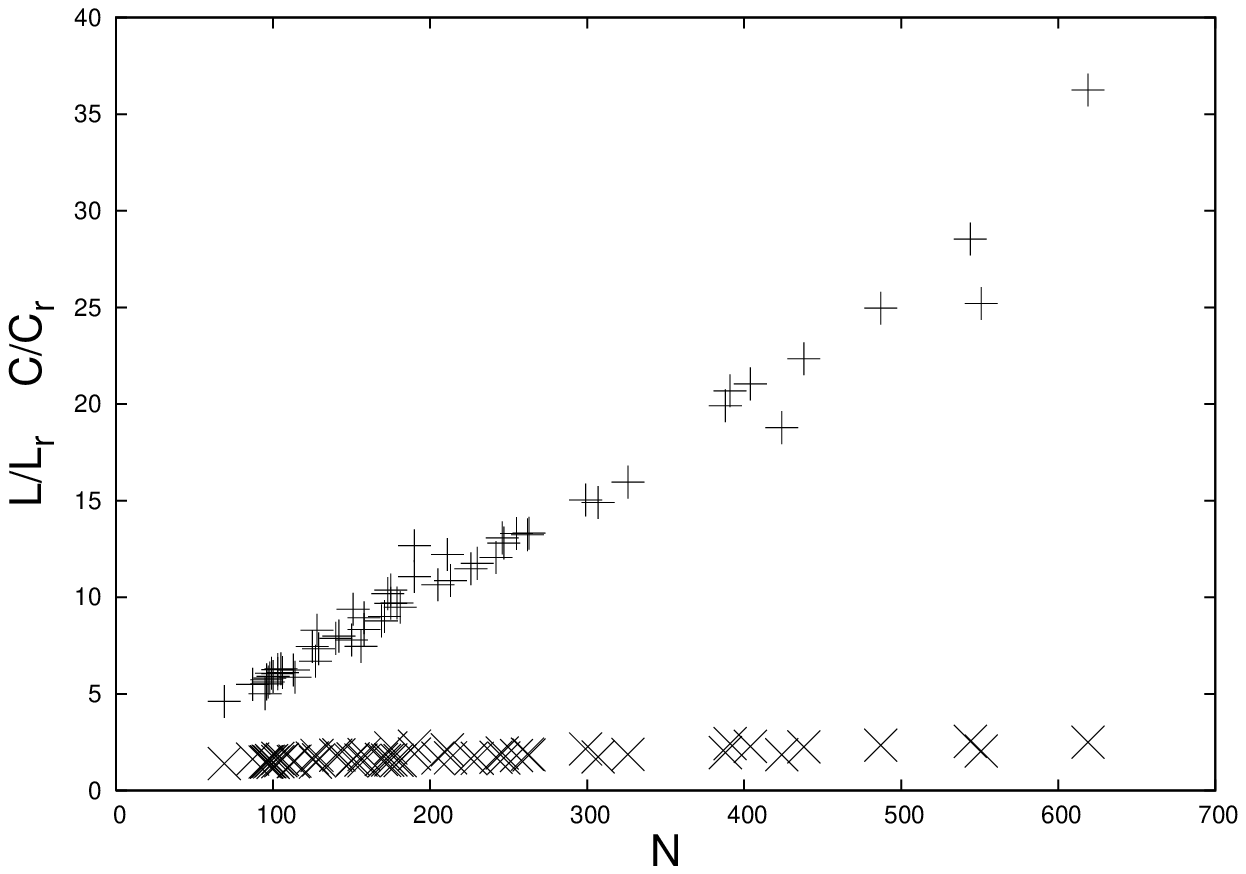}
\caption{The ratios p(=$<C>/<C_r>$) and q (=$<L>/<L_r>$) as a function of 
network size N. The ratio p (shown by plus sign) varies linearly with N; 
whereas the ratio q (shown by cross sign) varies logarothmically with N.}
\end{figure}
\begin{figure}
\includegraphics{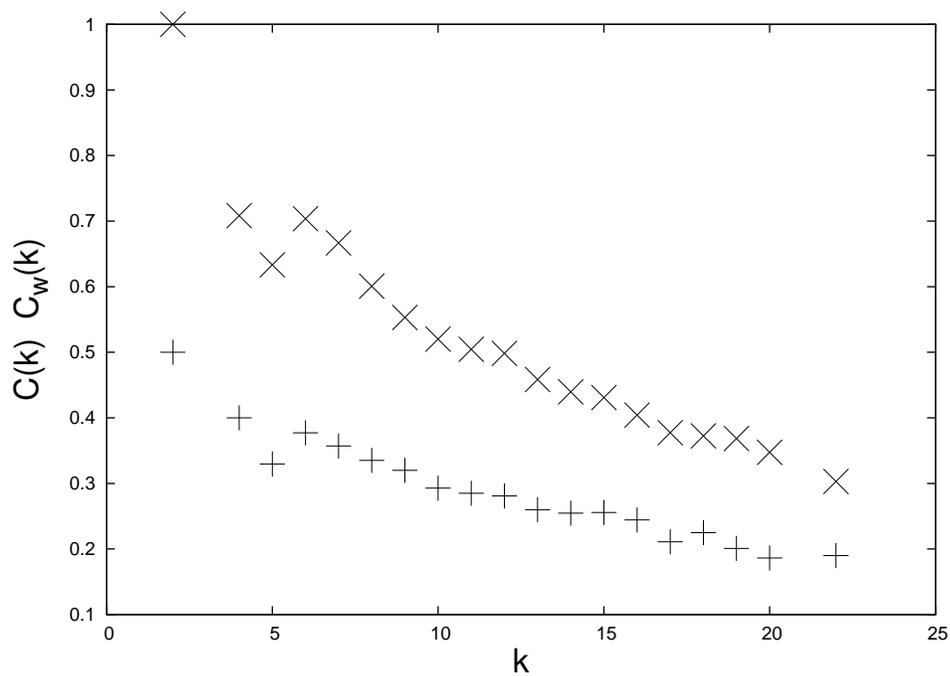}
\caption{A representative figure (PDB Id:1CD8) of topological clustering
coefficient C(k) (shown by cross sign) and weighted clustering coefficient
$C_w(k)$ (shown by plus sign) as a function of degree k of nodes.}
\end{figure}
\begin{figure}
\includegraphics{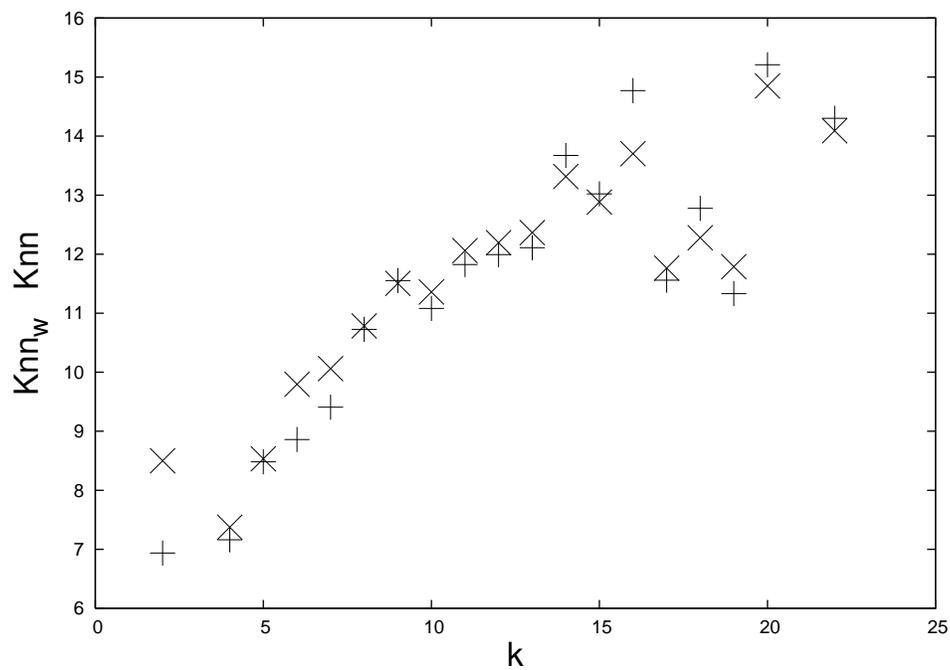}
\caption{A represented figure (PDB Id:1CD8) of unweighted $K_{nn}$ (shown 
by cross sign) and weighted $K_{nn}(w)$ (shown by plus sign) average 
degree of nearest neighbor as a function k. }
\end{figure}

\end{document}